\documentclass[times,twocolumn,final]{elsarticle}
\usepackage{medima}
\usepackage{framed,multirow}

\usepackage{amssymb}
\usepackage{latexsym}

\usepackage{url}
\usepackage{xcolor}

\usepackage[colorlinks, urlcolor=black, linkcolor=blue, citecolor=blue]{hyperref}

\definecolor{newcolor}{rgb}{.8,.349,.1}

\journal{Medical Image Analysis}

\usepackage{amsmath}
\usepackage{array}
\usepackage{booktabs}
\usepackage{multirow}
\usepackage{color}
\usepackage{colortbl}
\definecolor{R}{rgb}{1,0,0}
\definecolor{B}{rgb}{0,0,1}
\definecolor{g}{gray}{0.95}

\begin{document}

\verso{H. Pan \textit{et~al.}}

\begin{frontmatter}

\title{Breast Ultrasound Tumor Generation via Mask Generator and Text-Guided Network:\\A Clinically Controllable Framework with Downstream Evaluation}

\author[1,2]{Haoyu \snm{Pan}}
\author[3]{Hongxin \snm{Lin}}
\author[1,2]{Zetian \snm{Feng}}
\author[4]{Chuxuan \snm{Lin}}
\author[1]{Junyang \snm{Mo}}
\author[1]{Chu \snm{Zhang}}
\author[1]{Zijian \snm{Wu}}
\author[1,2]{Yi \snm{Wang}\corref{cor1}}
\author[5]{Qingqing \snm{Zheng}\corref{cor1}}
\cortext[cor1]{Corresponding authors: Yi Wang and Qingqing Zheng (emails: onewang@szu.edu.cn, zhengqingqing@suat-sz.edu.cn)}

\address[1]{School of Biomedical Engineering, Shenzhen University Medical School, Shenzhen University, Shenzhen, China}
\address[2]{Smart Medical Imaging, Learning and Engineering (SMILE) Lab, Shenzhen University, Shenzhen, China}
\address[3]{Nuclear Medicine Department, the Seventh Affiliated Hospital, Sun Yat-Sen University, Shenzhen, China}
\address[4]{Department of Radiology, the Seventh Affiliated Hospital, Sun Yat-Sen University, Shenzhen, China}
\address[5]{Shenzhen University of Advanced Technology, Shenzhen, China}

\received{June 2025}
\communicated{Y. Wang}

\begin{abstract}
The development of robust deep learning models for breast ultrasound (BUS) image analysis is significantly constrained by the scarcity of expert-annotated data.
To address this limitation, we propose a clinically controllable generative framework for synthesizing BUS images.
This framework integrates clinical descriptions with structural masks to generate tumors, enabling fine-grained control over tumor characteristics such as morphology, echogencity, and shape.
Furthermore, we design a semantic-curvature mask generator, which synthesizes structurally diverse tumor masks guided by clinical priors.
During inference, synthetic tumor masks serve as input to the generative framework, producing highly personalized synthetic BUS images with tumors that reflect real-world morphological diversity.
Quantitative evaluations on six public BUS datasets demonstrate the significant clinical utility of our synthetic images, showing their effectiveness in enhancing downstream breast cancer diagnosis tasks.
Furthermore, visual Turing tests conducted by experienced sonographers confirm the realism of the generated images, indicating the framework's potential to support broader clinical applications.
\emph{The code is publicly available.}
\end{abstract}

\begin{keyword}
\KWD 
\\Breast ultrasound\\
Diffusion model\\
Controllable generation\\
Data augmentation\\
Breast cancer diagnosis
\end{keyword}
\end{frontmatter}


\section{Introduction}
\label{sec:introduction}
Breast cancer is one of the most common types of cancer worldwide and early detection is key to improving patient survival rates~\citep{kim2025global}.
As a radiation-free and real-time imaging approach, ultrasound (US) has long been a primary tool for the screening and diagnosis of breast cancer~\citep{guo2018ultrasound}.
In recent years, deep learning has emerged as a pivotal technology in computer-aided diagnosis (CAD) systems for breast ultrasound (BUS), enabling automated tumor detection, classification, and segmentation~\citep{qian2021prospective, wang2019deeply, aljuaid2022computer, cho2025attention}.
However, the development of robust and generalizable deep learning models for BUS is hindered by the limited availability of annotated datasets, due to the high cost of expert labeling and data privacy concerns.

To mitigate data scarcity, there has been an increasing interest in the use of deep learning-based models for the generation of synthetic data in medical imaging.
Generative models, such as generative adversarial networks (GANs)~\citep{Goodfellow2014}, have demonstrated the ability to generate synthetic medical images in different modalities~\citep{han2022gan,liang2022sketch,dar2019image}.
These models offer a promising solution to augment limited annotated datasets.
However, GANs face challenges, including issues such as poor generation quality and training collapse~\citep{dhariwal2021diffusion}, where the model generates limited or repetitive patterns instead of diverse and accurate images.
Furthermore, GANs lack fine-grained controllability over clinically important features, such as tumor shape and internal texture, which are essential diagnostic indicators in BUS to distinguish between benign and malignant lesions.

Recently, the advent of diffusion models (DMs) has marked a notable shift in generative modeling, with increasing evidence demonstrating their superiority over GANs in terms of stability and visual fidelity.
Although some studies have extended DM to generation guided by structure or text conditioning, they are largely limited to general image domains and lack clinical relevance.
In BUS, challenges such as speckle noise, low contrast, and highly variable but clinically significant tumor appearances demand more specialized solutions. 
Existing methods either provide only coarse control over the generation process or lack guidance from clinical prior knowledge, limiting their ability in medical scenarios.
Specifically, clinically relevant features such as tumor morphology and echogenicity are key indicators for differentiating benign from malignant lesions, requiring fine-grained control during the generation process.
However, such clinically informed controllability remains largely underexplored in current diffusion-based frameworks.

\begin{figure}[t]
	\centering
	\includegraphics[width=\linewidth]{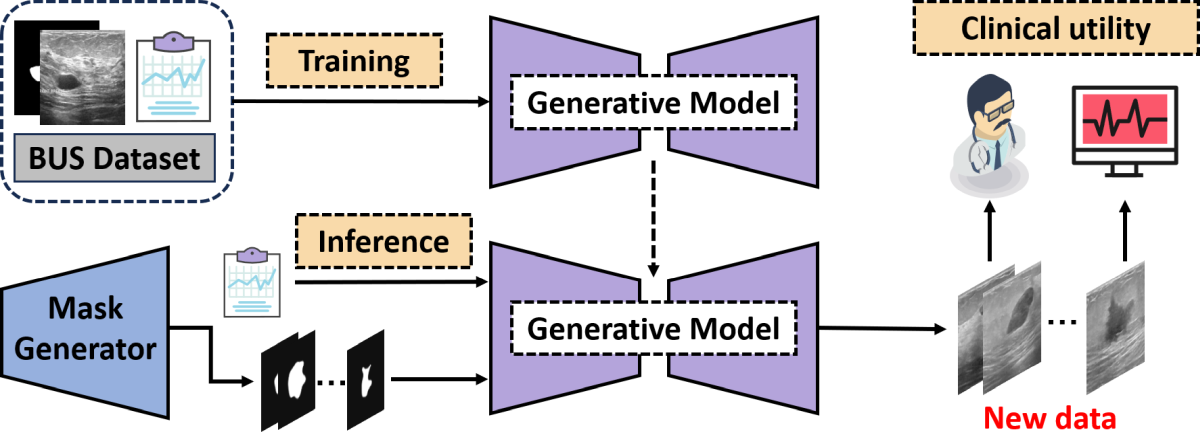}
	\caption{Overview of the proposed clinically controllable framework.}
	\label{fig:overview}
\end{figure}

To address these gaps, we propose a clinically controllable diffusion framework for BUS, which enables fine-grained control over tumor structure and texture, as shown in Fig.~\ref{fig:overview}.
Our framework integrates ControlNet with a denoising diffusion model to refine BUS generation. Built on the stable diffusion framework, our architecture incorporates prior knowledge via: 1) a \textbf{ControlNet module} that provides structural guidance using tumor masks; 
2) a \textbf{pre-trained text encoder} for clinical feature specification via prompts (e.g. echogenicity and boundary clarity);
and 3) a dedicated \textbf{shape-aware mask generator} that generates structurally diverse tumor masks based on tumor type (clinical label), curvature (shape prior), and bounding rectangle (location constraint). 
This design facilitates the generation of highly personalized synthetic tumors that reflect real-world morphological diversity. The resulting synthetic images can serve as a valuable data augmentation source to enhance the performance of CAD systems for breast cancer.

In summary, the contributions of this work are as follows:
\begin{itemize}
	\item We propose a clinically controllable BUS image generation framework that enables fine-grained control over tumor morphology and echogencity by integrating text and mask guidance.
	\item We design a shape-aware mask generator that synthesizes realistic and diverse tumor masks conditioned on interpretable structural descriptors, ensuring morphological plausibility and clinical relevance.
	\item The proposed framework generates clinically compliant synthetic data without the need for manual annotations, enabling effective enhancement of downstream classification and segmentation tasks. 
	\item We conduct extensive experiments on six BUS datasets, validating the effectiveness and generalizability of the proposed approach in multiple clinical scenarios and real-world diagnostic tasks.
\end{itemize}

\section{Related Work}
\label{r_w}
\subsection{Generative Models in Medical Image Synthesis}
Deep generative models, particularly GANs~\citep{Goodfellow2014} and DMs~\citep{Sohl-Dickstein;2015}, 
have revolutionized natural image synthesis and extended to medical imaging~\citep{kazerouni;2023}. 

Although GAN-based models have shown notable promise in various medical imaging modalities, including US~\citep{liang2022sketch}, X-ray~\citep{han2022gan}, computed tomography (CT)~\citep{jiang2020covid}, and magnetic resonance imaging (MRI)~\citep{dar2019image},
they often encounter challenges in unrealistic outputs and poor generalization~\citep{Thanh;2020}.
In contrast, the recent emergence of DMs, particularly the denoising diffusion probabilistic model (DDPM)~\citep{Ho;2020}, denoising diffusion implicit model (DDIM)~\citep{song2020denoising}, and the latent diffusion model (LDM)~\citep{Rombach;2022}, has further enhanced the quality of synthesized images.
These models have shown substantial improvements in training stability and are capable of generating high-quality images with rich details, making them effective for various applications in medical image generation.

\begin{figure*}[t]
	\centering
	\includegraphics[width=\linewidth]{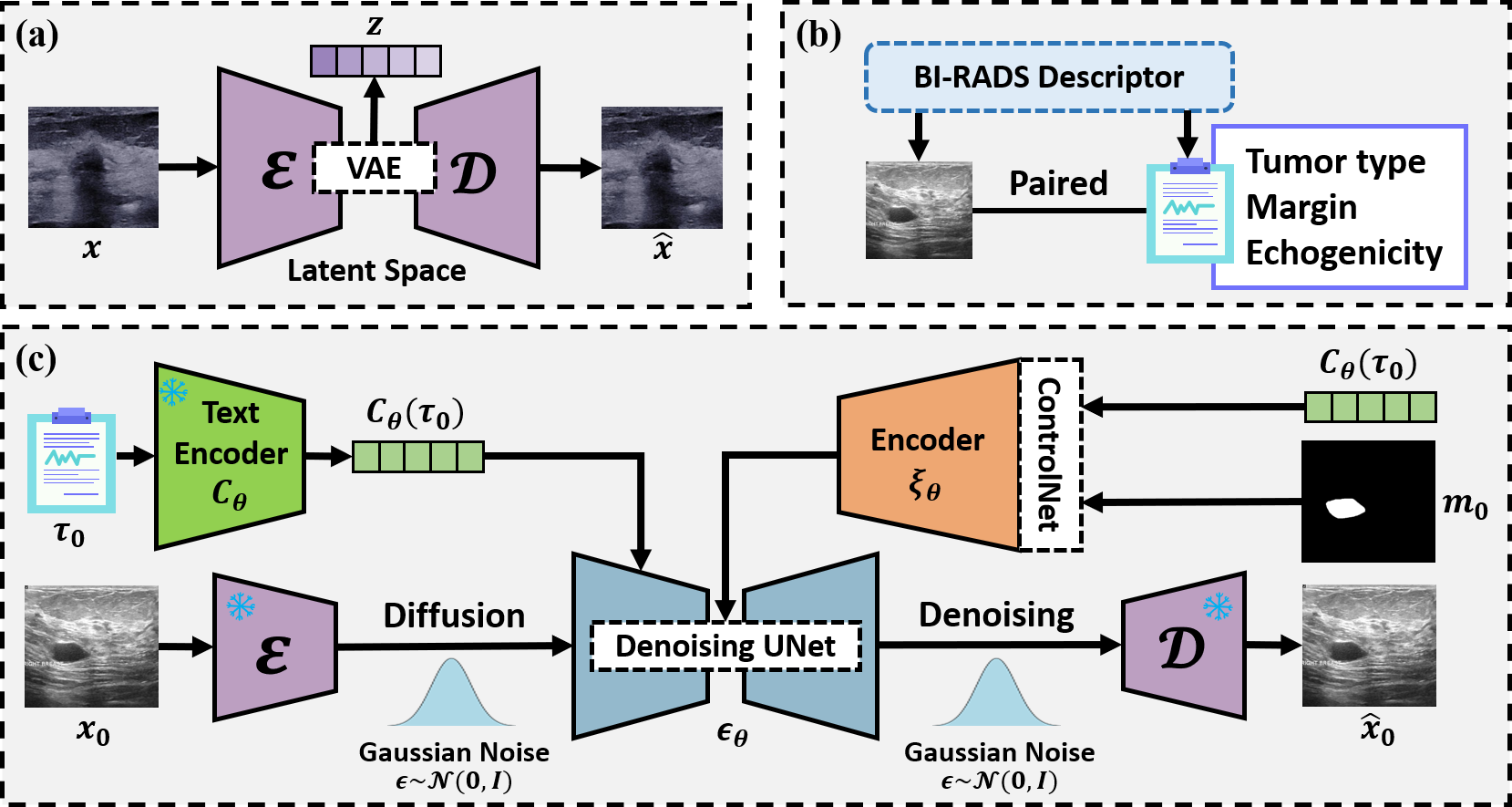}
	\caption{The training pipeline for generating breast ultrasound images with tumors follows a three-stage process: (a) the variational autoencoder (VAE) training, (b) the image-text pair design, and (c) the controlled diffusion process using the ControlNet framework.}
	\label{fig:training}
\end{figure*}

\subsection{Conditional Diffusion Models}
Conditional DMs introduce additional guidance mechanisms, such as text prompts and spatially localized input conditions, to regulate both semantic content and structural details~\citep{guo2024randomness}.
Therefore, controlling the generative process in DMs is crucial for clinical relevance.
Text-guided DMs~\citep{wang2025self} encode semantic features, including tumor type and texture pattern, but lack precise spatial control.  
To address above issues, researchers have explored incorporating spatial constraints such as masks~\citep{konz2024anatomically}, edge maps~\citep{zhang2024diffboost}, and sketches~\citep{koley2024s}.
Sketch-guided DMs generate images through abstract shape guidance, edge-guided DMs utilize contour information to refine anatomical structures, and mask-guided DMs provide explicit spatial control over specific organs or tumor regions. 
While these methods improve structural fidelity, they still fall short to capture diagnostic semantics, such as echogenicity and tumor type. 

More recently, hybrid approaches such as ControlNet~\citep{zhang2023adding} combining both text and spatial constraints (primarily mask-based conditioning) have been utilized in medical domains.
Such dual-modality conditioning enables image generation with fine-grained anatomical details, such as in BUS~\citep{freiche2025ultrasound}, colon polyp~\citep{sharma2024controlpolypnet}, and multi-modality brain MRI~\citep{Kim_2024_WACV}.
However, reliance on manual masks at inference limits scalability for variable tumor morphologies.

\subsection{Image Synthesis for Data Augmentation}
Synthetic data has been shown to enhance the performance of downstream models by serving as the augmented training data~\citep{goceri2023medical}.
For example, in the task of disease classification~\citep{luo2024measurement}, synthetic images have been shown to provide additional training data, and therefore reducing model overfitting and boosting classification accuracy. 
Similarly, in the task of lesion segmentation~\citep{fernandez2022can}, the inclusion of synthetic data has resulted in a more accurate boundary delineation, enabling segmentation models to handle more complex and diverse cases.

\section{Methodology}
\subsection{Clinically Controllable BUS Synthesis Framework}
\label{pipeline}
As illustrated in Fig.~\ref{fig:training}, our clinically controllable BUS synthesis framework integrates three key components:
(a) a variational autoencoder (VAE) that projects ultrasound images into a compressed latent space for efficient processing;
(b) a BI-RADS-aligned descriptor that extracts clinical descriptions (e.g., echogenicity, margin characteristics), and
(c) a dual-guidance generation module combining ControlNet for structural conditioning and text embeddings for semantic refinement within a stable diffusion backbone.
This unified architecture enables precise tumor morphology and tissue texture during BUS image synthesis.

\begin{figure*}[t]
	\centering
	\includegraphics[width=\linewidth]{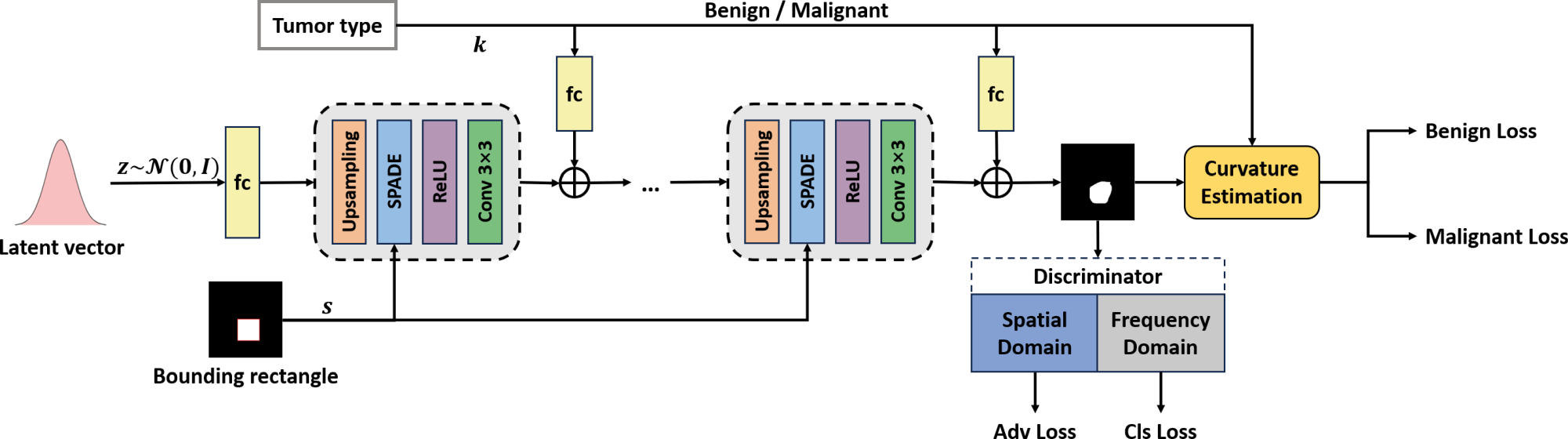}
	\caption{SCMG architecture overview. SCMG is a conditional generator that takes as input a latent vector $z$, a tumor type condition $k$, and a bounding rectangle $s$ indicating the tumor location and size. SCMG adopts a hierarchical upsampling structure where SPADE normalization blocks incorporate spatial control via the bounding rectangle. Semantic control is achieved by encoding the tumor type condition. A dual-branch discriminator evaluates both spatial realism and frequency consistency.}
	\label{fig:mask}
\end{figure*}

\subsubsection{Latent Space Construction}
To enable latent space-based diffusion modeling, we first train a VAE to project BUS images into a compact and semantically meaningful latent space. This latent representation serves as the input to our diffusion model, offering a balance between computational efficiency and structural fidelity.

The VAE comprises an encoder $\mathcal{E}$ and a decoder $\mathcal{D}$. Given a BUS image $x \in \mathbb{R}^{H \times W \times C}$ ($H$, $W$, $C$ denote height, width and channel number), the encoder $\mathcal{E}$ maps it to a latent distribution $q(z|x)$ parameterized by a mean $\mu(x)$ and variance $\sigma(x)^2$:
\begin{equation}
q(z|x) = \mathcal{N}(z \mid \mu(x), \sigma(x)^2).
\end{equation}
A latent vector $z$ is sampled using the reparameterization trick:
\begin{equation}
z = \mu(x) + \sigma(x) \cdot \epsilon, \quad \epsilon \sim \mathcal{N}(0, \mathbf{I}).
\end{equation}
The decoder $\mathcal{D}$ reconstructs the image from the latent vector as:
\begin{equation}
\hat{x} = \mathcal{D}(z).
\end{equation}

The training objective consists of a reconstruction loss and a regularization term. The reconstruction loss ensures fidelity between the input and output images:
\begin{equation}
\mathcal{L}_{\text{recon}} = \mathbb{E}_{q(z|x)} \left[ \| x - \hat{x} \|^2 \right].
\end{equation}
The Kullback-Leibler (KL) divergence regularizes the latent distribution toward a standard Gaussian prior:
\begin{equation}
\mathcal{L}_{\text{KL}} = \frac{1}{2} \sum \left( \mu(x)^2 + \sigma(x)^2 - \log \sigma(x)^2 - 1 \right).
\end{equation}
The total VAE loss function is the sum of the reconstruction loss and the KL divergence:
\begin{equation}
\mathcal{L}_{\text{VAE}} = \mathcal{L}_{\text{recon}} + \mathcal{L}_{\text{KL}}.
\end{equation}
The total loss ensures that the VAE both reconstructs the input images well and learns a smooth, structured latent space.

Once trained, the VAE parameters are frozen and reused throughout the diffusion model training. Specifically, the encoder $\mathcal{E}$ maps real images into the latent space to serve as ground truth targets, while the decoder $\mathcal{D}$ reconstructs images from the denoised latent vectors at the end of the diffusion process, see Fig.~\ref{fig:training}(c).
This latent-space formulation not only reduces computational complexity but also stabilizes diffusion training by filtering out low-level imaging noise. 

\subsubsection{BI-RADS-Guided Text Prompt Design}
We use the breast imaging reporting and data system (BI-RADS) to generate clinically precise text descriptions for paired BUS images.
The BI-RADS provides a structured method for categorizing tumor characteristics, focusing on attributes such as tumor \textbf{ benign or malignant nature}, \textbf{internal echogenicity}, and \textbf{border clarity}. 
Note that shape information is excluded from text prompts, as this is controlled through the tumor masks. This separation helps prevent conflicting guidance between textual and structural inputs, and avoid negative impact of redundant textual information.
Therefore, we tailor two specific templates to encode the distinctive characteristics between benign and malignant tumors.
For example,
``\textit{benign tumor} with well-defined borders and homogeneous internal echogenicity'', and 
``\textit{malignant tumor} with irregular borders and heterogeneous internal echogenicity''.

\subsubsection{BUS Synthesis with Structural and Textural Guidance} 
Given an input image \( x_0 \), its mask \( m_0 \), and the clinical description $\tau_0$, we first encode \( x_0 \) using the pre-trained VAE to obtain a latent representation \( z_0 \). The latent variable \( z_0 \) serves as the starting point for the forward diffusion process, where the noise is iteratively added over \( t \) timesteps to produce a noisy latent representation \( z_t \) with
\begin{equation}
z_t = \sqrt{\bar{\alpha}_t} \, z_0 + \sqrt{1 - \bar{\alpha}_t} \, \epsilon, 
\end{equation}
where $\bar{\alpha}_t$ is the cumulative product of noise schedule parameters. The denoising UNet $\epsilon_\theta$ is trained to reverse the diffusion process by predicting and removing noise from \( z_t \) in each timestep. This denoising process is conditioned on both textural and structural inputs. Specifically, $\tau_0$ is processed through the CLIP~\citep{radford2021learning} text encoder $\mathcal{C}_\theta$ to generate text embedding $\mathcal{C}_\theta(\tau_0)$. Meanwhile, $\mathcal{C}_\theta(\tau_0)$ and \( m_0 \) are processed by ControlNet $\xi_\theta$, generating the structural features $\xi_\theta(z_t,m_0,\tau_0)$. Both $\xi_\theta(z_t,m_0,\tau_0)$ and $\mathcal{C}_\theta(\tau_0)$ are subsequently incorporated into the denoising UNet as a residual feature. Thus, the denoising output is represented as $\epsilon_\theta(z_t, t, \mathcal{C}_\theta, \xi_\theta)$. The training objective employs the mean squared error (MSE) between the predicted and true noise, ensuring effective denoising:
\begin{equation}
\mathcal{L}_{\text{MSE}} = \mathbb{E}\left[\| \epsilon - \epsilon_\theta(z_t, t, \mathcal{C}_\theta, \xi_\theta) \|^2 \right].
\end{equation}

\subsection{Clinically Guided Tumor Mask Generation}
During inference, our method requires a tumor mask along with a BI-RADS-based text prompt to guide the diffusion process toward realistic and diagnostically meaningful tumor appearances.
To enable automatic mask generation, we propose the Semantic-Curvature Mask Generator (SCMG), a shape-aware module that synthesizes tumor masks using three clinical priors: tumor type (benign or malignant) for semantic control, bounding rectangle for spatial localization, and curvature regularization for boundary plausibility. As shown in Fig.~\ref{fig:mask}, SCMG eliminates manual annotation while ensuring diagnostic relevance through anatomically plausible mask synthesis.

\subsubsection{SCMG Design} 
The SCMG synthesizes tumor mask
\(\hat{M} = G(z, k, s)\)
through a hierarchical architecture that integrates clinical priors.
As shown in Fig.~\ref{fig:mask}, the generator processes three inputs, including the latent vector \( z \in \mathbb{R}^n \), the controllable tumor type \( k \), and the spatial guidance map (i.e., bounding rectangle \( s \)). The latent vector \( z \) is projected through a fully connected layer and reshaped into an initial feature map \( F_0 \in \mathbb{R}^{C_0 \times H_0 \times W_0} \). The tumor type \( k \) is embedded into a vector and spatially broadcast to match the dimensions of \( F_0 \), then concatenated with the feature map channel-wisely to maintain semantic consistency across scales. 
The feature map \( F_0 \) is processed through \(N\) sequential blocks, each performing: 
\begin{itemize}
	\item Nearest-neighbor upsampling to double the spatial resolution.
	\item A single SPADE~\citep{park2019semantic} normalization layer modulated by the resized spatial map \( s_n \):
	\begin{equation}
	\text{SPADE}(F_n, s_n) = \gamma(s_n) \cdot \text{IN}(F_n) + \beta(s_n),
	\end{equation}
	where \( \text{IN}(\cdot) \) denotes instance normalization, and \( \gamma(s_n) \) and \( \beta(s_n) \) are learned affine parameters.  
	\item A ReLU activation followed by \(3 \times 3\) convolution.
\end{itemize}

The final feature map undergoes \(3 \times 3\) convolution followed by sigmoid activation, producing a binary tumor mask with controllable type (benign or malignant), flexible positioning, and clinical plausibility of tumor morphology.

\subsubsection{Curvature-Aware Clinical Prior}
Benign tumors are typically characterized by smooth and regular boundaries, whereas malignant tumors often exhibit irregular and spiculated contours.
To explicitly model this clinical observation, we introduce a curvature-based regularization.

For each binary segmentation mask $M$, we extract the boundary pixel set $\mathcal{B}$ and compute the local curvature at each point using discrete differential operators.
The mean absolute curvature $\bar{\kappa}$ over the boundary is then defined as:
\begin{equation}
\bar{\kappa} = \frac{1}{N_b} \sum_{(i, j) \in \mathcal{B}} 
\left| \frac{M_{ii} M_j^2 - 2 M_i M_j M_{ij} + M_{jj} M_i^2}
{\left(M_i^2 + M_j^2 + \varepsilon\right)^{3/2}} \right|.
\label{eq:mean_absolute_curvature}
\end{equation}
Here, $N_b = |\mathcal{B}|$ denotes the total number of boundary pixels.
$M_i$, $M_j$, $M_{ii}$, $M_{jj}$, and $M_{ij}$ denote the first/second-order spatial derivatives along the horizontal and vertical axes.
$\varepsilon = 10^{-8}$ is for numerical stability. 
This formulation yields a differentiable, geometry-aware measure of contour irregularity, providing a quantitative basis for characterizing tumor boundary morphology.

Statistical analysis on real tumor masks reveals that benign tumors exhibit significantly lower curvature values ($0.679\pm0.151$) compared to malignant tumors ($0.789\pm0.163$), with a $p$-value $< 0.01$.
This motivates the design of curvature loss:
\begin{equation}
\mathcal{L}_{\text{cur}}  = \left\| \bar{\kappa}_{gen} - \bar{\kappa}_{target} \right\|,
\label{eq:loss_curvature}
\end{equation}
that encourages the curvature of generated masks ($\bar{\kappa}_{gen}$) to follow class-specific curvature distributions ($\bar{\kappa}_{target}$), thereby enforcing structural clinical priors.

\begin{table}[t]
	\centering
	\caption{Overview of the six public breast ultrasound datasets.}
	\label{tab:dataset}
	\resizebox{\columnwidth}{!}{
		\begin{tabular}{lcccc}
			\hline
			Dataset        & Images & Benign/Malignant    & Reference                      \\ \hline
			BUSI           & 647    & 437/210             &\cite{al2020dataset}            \\
			UDIAT          & 163    & 110/53              &\cite{yap2017automated}         \\
			BrEaST         & 252    & 154/98              &\cite{pawlowska2024curated}     \\
			BUS-UCLM       & 264    & 174/90              &\cite{vallez2025bus}            \\
			QAMEBI         & 232    & 109/123             &\cite{ardakani2023open}                   \\
			STU            & 42     & -                   &\cite{zhuang2019rdau}           \\ \hline
			\textbf{Total} & 1600   & 984/574            & -                              \\ \hline
	\end{tabular}}
\end{table}

\subsubsection{Adversarial Training with Clinical Priors}
To ensure anatomical realism and diagnostic relevance, we design a dual-branch discriminator \( D \), comprising parallel spatial and frequency branches.
The spatial branch employs stacked convolutional layers with instance normalization to analyze structural features, while the frequency branch computes Fourier magnitude spectra and convolutional encoding to assess boundary sharpness.
The proposed SCMG is optimized using a hybrid loss function that integrates four components:

\textbf{Adversarial Loss} ($\mathcal{L}_{\text{adv}}^{\text{D}}$, $\mathcal{L}_{\text{adv}}^{\text{G}}$):
Encourages the generator to produce masks indistinguishable from real samples by the discriminator:
\begin{equation}
\mathcal{L}_{\text{adv}}^{\text{D}} = \frac{1}{2} \left[ -\log D_{\text{adv}}(M) - \log \left(1 - D_{\text{adv}}(\hat{M}) \right) \right],
\end{equation}
\begin{equation}
\mathcal{L}_{\text{adv}}^{\text{G}} =  -\log D_{\text{adv}}(\hat{M}),
\end{equation}
where $D_{\text{adv}}$ denotes the real/fake probability.

\textbf{Classification Loss} ($\mathcal{L}_{\text{cls}}^{\text{D}}$):
Guides the discriminator to correctly identify tumor type based on frequency-domain cues:
\begin{equation}
\mathcal{L}_{\text{cls}}^{\text{D}} = -\left[ k \cdot \log D_{\text{cls}}(\hat{M}) + (1 - k) \cdot \log \left(1 - D_{\text{cls}}(\hat{M}) \right) \right]
\end{equation}
where $D_{\text{cla}}$ denotes the tumor type prediction, and $k$ is the tumor type.

\textbf{Curvature Loss} ($\mathcal{L}_{\text{cur}}$): Penalizes deviation of generated mask curvature from target class-specific statistics, see (\ref{eq:loss_curvature}).

The final training objectives for the proposed SCMG are:
\begin{equation}
\mathcal{L}_{\text{G}} = \mathcal{L}_{\text{adv}}^{\text{G}} + \mathcal{L}_{\text{cur}},
\end{equation}
\begin{equation}
\mathcal{L}_{\text{D}} = \mathcal{L}_{\text{adv}}^{\text{D}} + \mathcal{L}_{\text{cls}}^{\text{D}}.
\end{equation}
This joint optimization ensures synthesized masks preserve clinically plausible morphology while respecting pathological curvature distributions.

\section{Experiments and Results}
\subsection{Datasets and Implementations}
\subsubsection{Datasets}
We conducted our experiments on six public BUS datasets, as summarized in Table~\ref{tab:dataset}.
These datasets collectively comprise a total of 1600 BUS images with 984 benign and 574 malignant lesions.
Notably, the STU dataset~\citep{zhuang2019rdau} provides only segmentation masks without tumor type annotations, thus served exclusively as an external test set for the downstream segmentation task.
All datasets were preprocessed with intensity normalization and resizing to $256\times256$.
We adopted an 80\%/20\% split within each dataset for training and testing, respectively.

\begin{table*}[t]
	\centering
	\caption{Generation performance comparison of different methods.}
	\label{tab:comparison}
	\footnotesize
	\begin{tabular}{lcccccccccc}
		\hline
		Dataset   & \multicolumn{2}{c}{BUSI} & \multicolumn{2}{c}{UDIAT} & \multicolumn{2}{c}{BrEaST} & \multicolumn{2}{c}{BUS-UCLM} & \multicolumn{2}{c}{QAMEBI} \\ \hline
		Method    & FID↓        & KID↓       & FID↓         & KID↓       & FID↓         & KID↓        & FID↓          & KID↓         & FID↓         & KID↓        \\ \hline
		pix2pixHD~\citep{wang2018high} & 12.961      & 0.045      & 8.932        & 0.031      & 12.942       & 0.041       & 19.051        & 0.071        & 13.161       & 0.058       \\
		SPADE~\citep{park2019semantic}     & 9.780       & 0.032      & 9.377        & 0.031      & 11.932       & 0.033       & 17.736        & 0.062        & 8.344        & 0.028       \\
		SGD-DDIM~\citep{konz2024anatomically}  & 9.081       & 0.030      & 12.662       & 0.051      & 12.464       & 0.038       & 23.838        & 0.099        & 10.862       & 0.042       \\
		SGD-DDPM~\citep{konz2024anatomically}  & 16.609      & 0.070      & 11.450       & 0.044      & 9.054        & 0.013       & 20.654        & 0.055        & 10.356       & 0.033       \\
		ArSDM~\citep{du2023arsdm}  & 8.864      & 0.023      & 9.189       & 0.017      & 10.392        & 0.015       & 17.380        & 0.039        & 11.692       & 0.024       \\
		Ours-t    & 14.735      & 0.032      & 8.984        & 0.025      & 12.806       & 0.027       & 21.096        & 0.064        & 11.651       & 0.029       \\
		Ours-m    & 7.319          & 0.019          & \textbf{6.525} & \textbf{0.008} & 11.798         & 0.021          & 18.936          & 0.047          & 8.750          & 0.021          \\
		Ours      & \textbf{6.770} & \textbf{0.013} & 7.072          & 0.013          & \textbf{8.881} & \textbf{0.018} & \textbf{16.614} & \textbf{0.036} & \textbf{6.543} & \textbf{0.010} \\ \hline
	\end{tabular} %
\end{table*}

\subsubsection{Implementations}
We adopted the pretrained VAE weights from stable diffusion~\citep{Rombach;2022} and fine-tuned them across training sets of all datasets.
This VAE was then integrated into our generation pipeline.
ControlNet~\citep{zhang2023adding} was trained for 200 epochs using gradient accumulation with an effective batch size of 128.
We utilized the AdamW optimizer with a learning rate of 1e-4.
The SCMG was trained with a batch size of 64.
The parameters of its generator and discriminator were optimized using Adam, with learning rates of 2e-4 and 1e-4, respectively.
For downstream tasks (i.e., classification and segmentation models), Adam optimizer was used with a learning rate of 1e-5, training with a batch size of 24. 

\subsubsection{Evaluation Metrics}
To evaluate the generation quality, we reported Fréchet Inception Distance (FID) and Kernel Inception Distance (KID).
For the downstream tasks,
classification was evaluated using the area under receiver operating characteristic curve (AUC) and F1-score,
while segmentation was measured using the Dice similarity coefficient (DSC).

\subsection{Comparison on BUS Tumor Generation}
\subsubsection{Quantitative Assessment}
We evaluated the proposed method against the state-of-the-art (SOTA) image synthesis approaches, including GAN-based and diffusion-based approaches.
Among GAN-based methods, we compared with pix2pixHD~\citep{wang2018high} and SPADE~\citep{park2019semantic}, both of which generate images from semantic masks.
For diffusion-based approaches, we included segmentation-guided diffusion (SGD)~\citep{konz2024anatomically}, a framework designed for anatomically controllable medical image generation that supports DDPM and DDIM sampling strategies.
Moreover, an adaptive refinement semantic diffusion models (ArSDM)~\citep{du2023arsdm} was also included for the comparison.

\begin{figure}[t]
	\centering
	\includegraphics[width=\columnwidth]{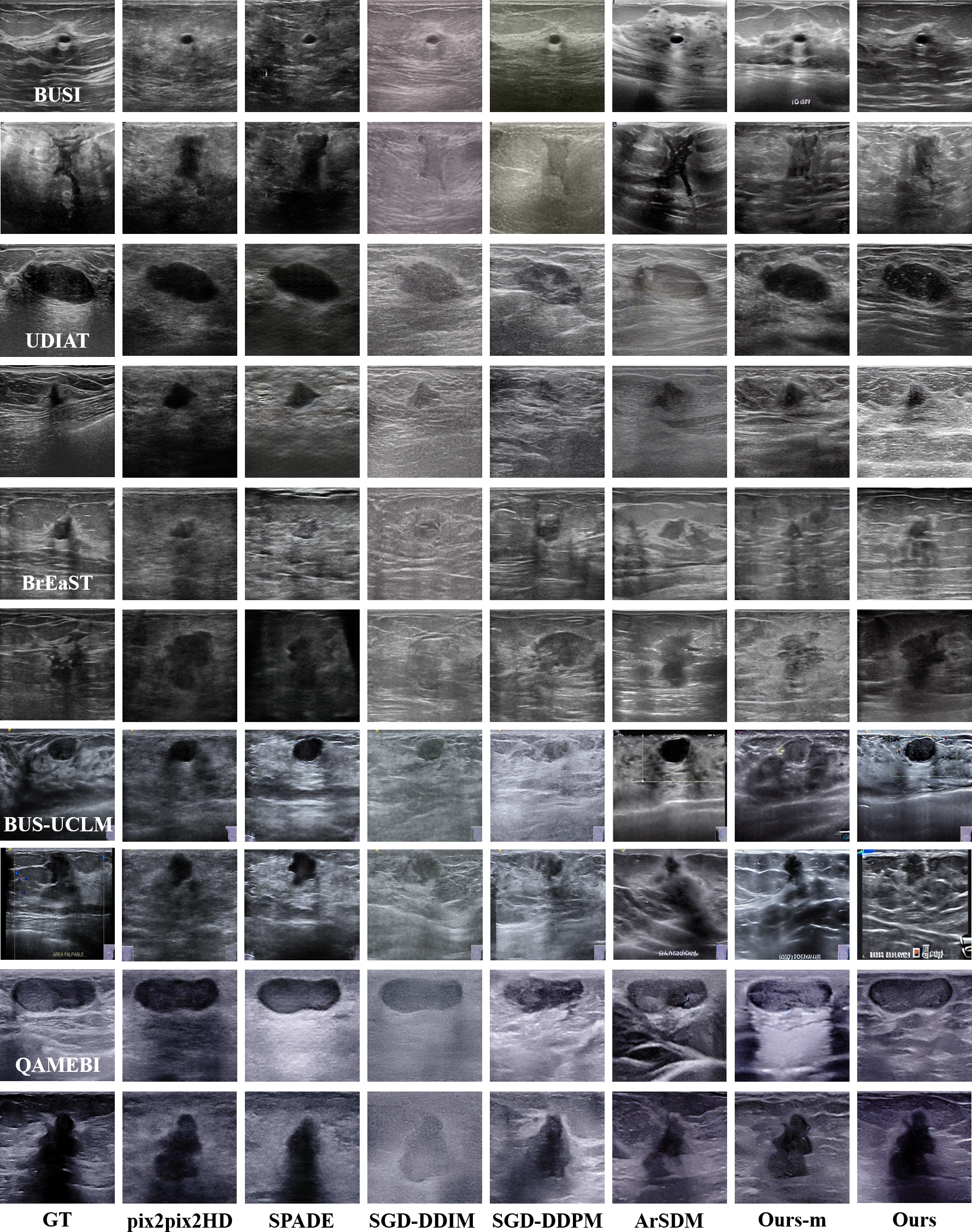}
	\caption{Visual comparison of breast ultrasound images generated by different methods. For each dataset, the top and bottom rows show benign and malignant cases, respectively.}
	\label{fig:results}
\end{figure}

Quantitative results are shown in Table~\ref{tab:comparison}.
Across all datasets, our method achieved overall the best performance, consistently outperforming baseline methods in both FID and KID.
Specifically, our full model combining text and mask guidance (denoted as \textbf{Ours}) achieved the lowest FID and KID on BUSI (FID=6.770, KID=0.013), BrEaST (FID=8.881, KID=0.018), BUS-UCLM (FID=16.614, KID=0.036), and QAMEBI (FID=6.543, KID=0.010).
On UDIAT, our mask-guided variant (Ours-m) achieved the best results (FID=6.525, KID=0.008), underscoring the effectiveness of explicit mask guidance.

Compared to GAN-based methods (pix2pixHD, SPADE), our approach demonstrates substantial improvements across all datasets, evidencing the advantage of combining diffusion-based generative models with clinically guided mask and text conditioning.
In particular, diffusion-based baselines (SGD-DDIM, SGD-DDPM, ArSDM), which also employed tumor masks, attained competitive results but still suffered performance degradation on certain datasets such as UDIAT, BUS-UCLM and QAMEBI even when compared to GAN-based methods.
This suggests that mask guidance alone may be insufficient to ensure consistent tumor structure generation, particularly under complex morphological variations.
The superior performance of our method highlights the benefit of integrating both mask and text guidance with clinical prior constraints to enhance controllability and fidelity.

Ablation results in Table~\ref{tab:comparison} further confirm the contribution of our design.
The text-guided only variant (Ours-t) underperformed comparing to the mask-guided (Ours-m) and full method (Ours), highlighting the critical role of structural control in preserving tumor morphology.
The consistent superiority of our full model across diverse datasets further demonstrates its robustness and clinical fidelity.

\begin{table}[t]
	\centering
	\caption{Visual Turing Test accuracy (\%) of three sonographers in distinguishing real and synthetic images generated by different methods. 
	}
	\label{tab:turing}
	\footnotesize
	\resizebox{\columnwidth}{!}{
		\begin{tabular}{llcccc}
			\hline
			Dataset                   & Method        & S1            & S2            & S3            & Average          \\ \hline
			\multirow{5}{*}{BUSI}     & pix2pixHD     & 69.0          & 74.0          & 70.0          & 71.0±2.2          \\
			& SPADE         & 71.0          & 79.0          & 72.0          & 74.0±3.6          \\
			& SGD-DDIM      & 80.0          & 85.0          & 89.0          & 84.7±3.7          \\
			& SGD-DDPM      & 84.0          & 85.0          & 80.0          & 83.0±2.2          \\
			& ArSDM         & 62.0          & 55.0          & 70.0          & 62.3±6.1          \\
			& Ours & \textbf{45.0} & \textbf{45.0} & \textbf{49.0} & \textbf{46.3±1.9} \\ \hline
			\multirow{5}{*}{UDIAT}    & pix2pixHD     & 76.0          & 80.0          & 81.0          & 79.0±2.2          \\
			& SPADE         & 75.0          & 79.0          & 74.0          & 76.0±2.2          \\
			& SGD-DDIM      & 82.0          & 86.0          & 88.0          & 85.3±2.5          \\
			& SGD-DDPM      & 78.0          & 71.0          & 76.0          & 75.0±2.9            \\
			& ArSDM         & 64.0          & \textbf{53.0}          & 60.0          & 59.0±4.5          \\
			& Ours & \textbf{51.0} & 44.0 & \textbf{51.0} & \textbf{48.7±3.3} \\ \hline
			\multirow{5}{*}{BrEaST}   & pix2pixHD     & 66.0          & 62.0          & 68.0          & 65.3±2.5          \\
			& SPADE         & 61.0          & 59.0          & 67.0          & 62.3±3.4          \\
			& SGD-DDIM      & 80.0          & 84.0          & 82.0          & 82.0±1.6          \\
			& SGD-DDPM      & 69.0          & 65.0          & 60.0          & 64.7±3.7          \\
			& ArSDM         & 54.0          & 66.0          & 58.0          & 59.3±5.0          \\
			& Ours & \textbf{47.0} & \textbf{45.0} & \textbf{50.0} & \textbf{47.3±2.1} \\ \hline
			\multirow{5}{*}{BUS-UCLM} & pix2pixHD     & 70.0          & 75.0          & 70.0          & 71.7±2.4          \\
			& SPADE         & 62.0          & 64.0          & 61.0          & 62.3±1.2          \\
			& SGD-DDIM      & 84.0          & 81.0          & 79.0          & 81.3±2.1          \\
			& SGD-DDPM      & 75.0          & 77.0          & 73.0          & 75.0±1.6          \\
			& ArSDM         & 72.0          & 63.0          & 61.0          & 65.3±4.8          \\
			& Ours & \textbf{61.0} & \textbf{56.0} & \textbf{55.0} & \textbf{57.3±2.6} \\ \hline
			\multirow{5}{*}{QAMEBI}   & pix2pixHD     & 69.0          & 71.0          & 65.0          & 68.3±2.5          \\
			& SPADE         & 72.0          & 69.0          & 64.0          & 68.3±3.3          \\
			& SGD-DDIM      & 90.0          & 85.0          & 88.0          & 87.7±2.1          \\
			& SGD-DDPM      & 70.0          & 69.0          & 64.0          & 67.7±2.6          \\
			& ArSDM         & 73.0          & 67.0          & 69.0          & 70.0±2.5          \\
			& Ours & \textbf{55.0} & \textbf{54.0} & \textbf{47.0} & \textbf{52.0±3.6} \\ \hline
		\end{tabular}
	}
\end{table}

\begin{table}[t]
	\centering
	\caption{Ablation of curvature loss: boundary curvature values (mean$\pm$std) for benign and malignant tumor masks.}
	\footnotesize
	\label{tab:curvature_ablation}
	\begin{tabular}{lcc}
		\toprule
		\textbf{Model Setting} & \textbf{Benign} & \textbf{Malignant} \\
		\midrule
		w/o $\mathcal{L}{\text{cur}}$ & 0.723$\pm$0.191 & 0.721$\pm$0.164 \\
		w/ $\mathcal{L}{\text{cur}}$ & 0.619$\pm$0.173 & 0.831$\pm$0.125 \\
		\bottomrule
	\end{tabular}
\end{table}

\begin{figure}[t]
	\centering
	\includegraphics[width=1\linewidth]{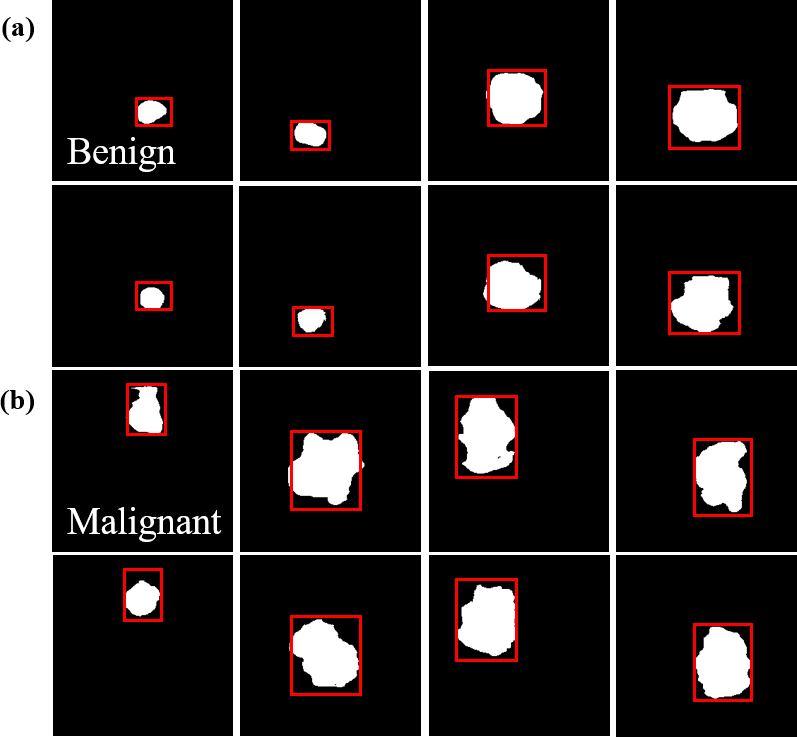}
	\caption{Visual comparison of generated tumor masks for (a) benign and (b) malignant settings.
		For each type, the top row uses curvature loss; the bottom row omits it. Target location and size are marked by the red input bounding rectangle.
	}
	\label{fig:mask results}
\end{figure}

\subsubsection{Qualitative Assessment}
Fig.~\ref{fig:results} presents visual comparison of BUS image generation results across different methods and datasets, including both benign and malignant cases. 

Overall, GAN-based methods provided reasonable control over tumor shape but struggled to accurately synthesize US-specific textures, often resulting in oversimplified lesion appearances and insufficient representation of acoustic features such as echogenicity and boundary details.
Among diffusion-based methods, SGD-DDIM had limited ability to preserve the intended tumor shape, producing blurry lesions with substantial variations in image tone compared to the original datasets.
SGD-DDPM and ArSDM achieved improved quality, however, the generated images still exhibited noticeable discrepancies in grayscale distribution and lacked the fine-grained texture characteristics of real US images.

In contrast, our method generated tumor images that not only accurately followed the prescribed shape guidance but also exhibited realistic ultrasound textures, boundary clarity, and echogenicity patterns consistent with the characteristics of each dataset. 
These observations highlight the superior capability of our framework in synthesizing clinically plausible and visually consistent BUS images across diverse domains.

\begin{table*}[t]
	\centering
	\caption{Classification performance (AUC/F1-score, mean±std) across datasets under different synthetic-to-real mixing ratios. Bold number denotes the best performance of each model.}
	\label{tab:cls-summary}
	\resizebox{\textwidth}{!}{
		\begin{tabular}{llcccccccccccc}
			\toprule
			Dataset & Model & \multicolumn{2}{c}{0\%} & \multicolumn{2}{c}{Ordinary Augmentation} & \multicolumn{2}{c}{25\%} & \multicolumn{2}{c}{50\%} & \multicolumn{2}{c}{100\%} & \multicolumn{2}{c}{200\%} \\
			\cmidrule(lr){3-14}
			& & AUC & F1-score & AUC & F1-score & AUC & F1-score & AUC & F1-score & AUC & F1-score & AUC & F1-score \\
			\midrule
			BUSI & ResNet50 & 0.843±0.013 & 0.779±0.022 & 0.857±0.011 & 0.796±0.015 & \textbf{0.882±0.007} & \textbf{0.835±0.012} & 0.860±0.010 & 0.807±0.014 & 0.873±0.010 & 0.827±0.010 & 0.864±0.004 & 0.814±0.005 \\
			& DenseNet121 & 0.871±0.005 & 0.822±0.010 & 0.889±0.008 & 0.846±0.011 & \textbf{0.908±0.004} & \textbf{0.870±0.005} & 0.896±0.007 & 0.854±0.010 & 0.901±0.011 & 0.858±0.013 & 0.908±0.004 & 0.868±0.007 \\
			& EfficientNet-B0 & 0.847±0.008 & 0.797±0.010 & 0.854±0.006 & 0.801±0.008 & 0.861±0.008 & 0.811±0.006 & \textbf{0.879±0.004} & 0.827±0.009 & 0.863±0.019 & 0.810±0.024 & 0.877±0.010 & \textbf{0.828±0.013} \\
			& MobileNetV2 & 0.826±0.018 & 0.764±0.023 & 0.842±0.011 & 0.779±0.014 & 0.853±0.012 & 0.790±0.016 & 0.863±0.026 & 0.820±0.004 & \textbf{0.873±0.005} & \textbf{0.823±0.005} & 0.856±0.014 & 0.780±0.016 \\
			\midrule
			UDIAT & ResNet50 & 0.688±0.021 & 0.608±0.058 & 0.695±0.024 & 0.619±0.038 & 0.726±0.032 & 0.684±0.037 & 0.688±0.015 & 0.604±0.032 & \textbf{0.750±0.011} & 0.689±0.031 & 0.746±0.021 & \textbf{0.717±0.026} \\
			& DenseNet121 & 0.667±0.021 & 0.612±0.109 & 0.677±0.014 & 0.616±0.069 & 0.697±0.008 & 0.690±0.040 & 0.692±0.011 & 0.648±0.044 & \textbf{0.728±0.009} & \textbf{0.726±0.002} & 0.699±0.010 & 0.694±0.041 \\
			& EfficientNet-B0 & 0.660±0.005 & 0.575±0.045 & 0.657±0.013 & 0.559±0.055 & \textbf{0.710±0.016} & \textbf{0.632±0.022} & 0.691±0.011 & 0.606±0.033 & 0.689±0.011 & 0.610±0.036 & 0.679±0.018 & 0.589±0.035 \\
			& MobileNetV2 & 0.679±0.037 & 0.592±0.038 & 0.684±0.029 & 0.588±0.042 & \textbf{0.733±0.027} & \textbf{0.726±0.019} & 0.731±0.012 & 0.698±0.030 & 0.721±0.021 & 0.676±0.046 & 0.699±0.039 & 0.678±0.040 \\
			\midrule
			BrEaST & ResNet50 & 0.741±0.034 & 0.694±0.034 & 0.743±0.019 & 0.670±0.022 & \textbf{0.784±0.027} & \textbf{0.741±0.028} & 0.752±0.019 & 0.670±0.024 & 0.744±0.014 & 0.695±0.016 & 0.745±0.011 & 0.690±0.015 \\
			& DenseNet121 & 0.759±0.011 & 0.710±0.014 & 0.763±0.014 & 0.719±0.025 & \textbf{0.789±0.003} & \textbf{0.743±0.005} & 0.766±0.014 & 0.718±0.015 & 0.782±0.011 & 0.731±0.014 & 0.757±0.023 & 0.699±0.026 \\
			& EfficientNet-B0 & 0.724±0.008 & 0.669±0.017 & 0.727±0.012 & 0.671±0.018 & 0.755±0.021 & 0.705±0.029 & 0.735±0.022 & 0.685±0.024 & \textbf{0.773±0.012} & \textbf{0.723±0.015} & 0.729±0.018 & 0.676±0.024 \\
			& MobileNetV2 & 0.684±0.016 & 0.593±0.047 & 0.708±0.026 & 0.607±0.027 & 0.719±0.003 & 0.658±0.007 & 0.738±0.005 & 0.687±0.013 & \textbf{0.757±0.022} & \textbf{0.703±0.026} & 0.727±0.005 & 0.673±0.006 \\
			\midrule
			BUS-UCLM & ResNet50 & 0.834±0.005 & 0.769±0.006 & 0.856±0.009 & 0.794±0.010 & \textbf{0.879±0.014} & \textbf{0.830±0.014} & 0.859±0.020 & 0.803±0.026 & 0.870±0.010 & 0.819±0.013 & 0.852±0.019 & 0.780±0.024 \\
			& DenseNet121 & 0.860±0.023 & 0.818±0.024 & 0.875±0.013 & 0.831±0.017 & 0.888±0.017 & 0.849±0.016 & \textbf{0.902±0.008} & \textbf{0.873±0.009} & 0.870±0.014 & 0.828±0.022 & 0.863±0.021 & 0.818±0.034 \\
			& EfficientNet-B0 & 0.800±0.039 & 0.730±0.046 & 0.818±0.020 & 0.753±0.025 & \textbf{0.847±0.018} & \textbf{0.779±0.021} & 0.826±0.024 & 0.769±0.026 & 0.807±0.020 & 0.739±0.021 & 0.824±0.006 & 0.761±0.009 \\
			& MobileNetV2 & 0.761±0.031 & 0.678±0.038 & 0.789±0.018 & 0.712±0.022 & \textbf{0.807±0.024} & 0.740±0.030 & 0.768±0.027 & 0.689±0.035 & 0.790±0.036 & \textbf{0.749±0.059} & 0.788±0.015 & 0.742±0.036 \\
			\midrule
			QAMEBI & ResNet50 & 0.755±0.009 & 0.788±0.010 & 0.764±0.008 & 0.775±0.012 & 0.775±0.007 & 0.767±0.013 & \textbf{0.796±0.004} & 0.790±0.021 & 0.787±0.007 & 0.773±0.027 & 0.786±0.016 & \textbf{0.802±0.011} \\
			& DenseNet121 & 0.798±0.012 & 0.798±0.014 & 0.807±0.010 & 0.826±0.013 & 0.816±0.009 & 0.836±0.012 & 0.826±0.010 & \textbf{0.844±0.023} & \textbf{0.828±0.009} & 0.841±0.022 & 0.810±0.027 & 0.812±0.008 \\
			& EfficientNet-B0 & 0.705±0.012 & 0.698±0.034 & 0.748±0.017 & 0.721±0.030 & 0.752±0.019 & 0.753±0.029 & \textbf{0.778±0.022} & \textbf{0.784±0.020} & 0.752±0.021 & 0.724±0.041 & 0.737±0.021 & 0.734±0.024 \\
			& MobileNetV2 & 0.724±0.021 & 0.730±0.032 & 0.752±0.018 & 0.759±0.028 & 0.770±0.026 & 0.783±0.039 & 0.752±0.011 & 0.753±0.010 & \textbf{0.783±0.018} & \textbf{0.784±0.038} & 0.755±0.017 & 0.764±0.039 \\
			\bottomrule
	\end{tabular}}
\end{table*}

\begin{table*}[t]
	\centering
	\caption{Segmentation performance (DSC\%, mean$\pm$std) across datasets under different synthetic-to-real mixing ratios. Bold number denotes the best performance of each model on the internal and external datasets.}
	\label{tab:seg-summary}
	\resizebox{\textwidth}{!}{
		\begin{tabular}{lllcccccccccccc}
			\toprule
			Dataset & Model & \multicolumn{2}{c}{0\%} & \multicolumn{2}{c}{Ordinary Augmentation} & \multicolumn{2}{c}{25\%} & \multicolumn{2}{c}{50\%} & \multicolumn{2}{c}{100\%} & \multicolumn{2}{c}{200\%} \\
			\cmidrule(lr){3-14}
			& & Internal & STU & Internal & STU & Internal & STU & Internal & STU & Internal & STU & Internal & STU \\
			\midrule
			BUSI & UNet & 79.6±0.3 & 82.6±0.2 & 79.8±0.2 & 82.8±0.4 & 80.9±0.3 & 83.2±0.3 & 80.5±0.2 & 82.7±0.6 & \textbf{81.5±0.4} & \textbf{87.0±0.5} & 81.0±0.6 & 85.5±0.3 \\
			& UNet++ & 77.3±0.6 & 84.3±0.5 & 77.8±0.5 & 84.6±0.3 & 79.0±0.5 & 88.4±0.4 & 79.8±0.4 & \textbf{88.5±0.2} & 80.4±0.5 & 86.6±0.4 & \textbf{80.9±1.0} & 85.0±0.6 \\
			& ResUNet & 79.2±0.5 & 85.1±0.2 & 79.5±0.3 & \textbf{85.3±0.4} & 80.0±0.2 & 84.9±0.5 & 80.1±0.8 & 85.2±0.3 & \textbf{80.3±0.2} & 84.9±0.3 & 80.3±0.5 & 84.8±0.4 \\
			& Attention UNet & 79.7±0.4 & 85.7±0.4 & 79.9±0.5 & 85.5±0.6 & 80.9±0.6 & 88.2±0.7 & \textbf{81.2±0.2} & 87.2±0.6 & 80.7±0.2 & \textbf{88.5±0.4} & 81.0±0.4 & 86.7±0.5 \\
			\midrule
			UDIAT & UNet & 83.6±0.4 & 56.6±1.7 & 82.9±0.4 & 57.3±0.9 & 83.4±0.2 & 72.8±0.3 & 83.8±0.5 & \textbf{80.6±0.2} & 84.6±0.5 & 78.8±0.9 & \textbf{84.9±0.5} & 75.5±0.3 \\
			& UNet++ & 82.5±0.2 & 52.5±1.0 & 82.7±0.4 & 60.4±0.5 & 83.4±0.3 & 70.6±0.9 & 84.3±0.5 & \textbf{80.1±0.8} & 85.0±0.5 & 79.3±0.5 & \textbf{85.6±0.8} & 76.9±0.2 \\
			& ResUNet & 86.9±0.5 & 54.3±0.9 & 86.7±0.4 & 55.7±0.4 & 86.4±0.5 & 73.4±0.4 & 86.6±0.7 & \textbf{80.6±0.1} & 88.2±0.4 & 79.5±0.2 & \textbf{88.3±0.5} & 76.8±0.4 \\
			& Attention UNet & 83.7±0.3 & 65.9±1.1 & 84.3±0.6 & 69.4±0.3 & 86.6±0.2 & 76.3±0.8 & 86.5±0.2 & 80.4±0.6 & 85.6±0.4 & 78.9±0.3 & \textbf{88.4±0.5} & \textbf{80.7±0.5} \\
			\midrule
			BrEaST & UNet & 67.7±0.3 & 60.3±0.5 & 69.8±0.4 & 67.4±0.2 & 70.4±0.2 & 72.2±0.3 & 72.4±0.6 & 76.1±0.3 & \textbf{73.3±0.5} & 81.6±0.3 & 72.4±0.3 & \textbf{81.9±0.5} \\
			& UNet++ & 69.3±0.5 & 58.4±0.5 & 69.5±0.2 & 60.5±0.6 & 68.0±0.7 & 70.6±0.3 & \textbf{71.9±0.1} & 78.6±0.5 & 70.3±0.3 & \textbf{81.1±0.1} & 70.8±0.3 & 78.2±0.2 \\
			& ResUNet & 74.0±0.2 & 56.4±0.3 & 73.6±0.5 & 59.4±0.7 & 74.9±0.5 & 73.9±0.3 & 76.6±0.2 & 75.0±0.2 & \textbf{76.7±0.3} & \textbf{83.3±0.2} & 74.5±0.4 & 83.1±0.1 \\
			& Attention UNet & 71.5±0.6 & 63.7±0.4 & 71.9±0.3 & 66.4±0.2 & 73.1±0.6 & 79.1±0.5 & 73.8±0.4 & 79.6±0.7 & 74.2±0.1 & \textbf{84.4±0.2} & \textbf{75.4±0.6} & 82.7±0.1 \\
			\midrule
			BUS-UCLM & UNet & 78.8±0.6 & 64.3±0.6 & 78.7±0.4 & 76.2±0.5 & 78.5±0.2 & 75.0±0.4 & 80.3±0.5 & 78.0±0.3 & \textbf{81.8±0.2} & \textbf{81.1±0.6} & 79.9±0.3 & 79.9±0.7 \\
			& ResUNet & 80.2±0.4 & 69.5±0.3 & 81.7±0.4 & 78.8±0.6 & \textbf{83.2±0.3} & 82.3±0.9 & 80.1±0.6 & \textbf{83.4±0.5} & 81.4±0.3 & 81.7±0.3 & 76.7±0.5 & 76.4±0.6 \\
			& UNet++ & 78.9±1.0 & 68.0±0.8 & 79.5±0.6 & 74.1±0.7 & 79.0±0.5 & 76.3±0.8 & 79.8±0.3 & 81.2±0.3 & \textbf{80.4±0.6} & \textbf{81.4±0.2} & 79.5±0.4 & 78.1±0.5 \\
			& Attention UNet & 81.5±0.5 & 71.1±0.4 & 81.9±0.3 & 73.9±0.5 & \textbf{83.5±0.6} & 80.0±0.6 & 82.6±0.5 & \textbf{82.5±0.4} & 80.9±0.6 & 81.1±0.5 & 82.7±0.3 & 80.2±0.6 \\
			\midrule
			QAMEBI & UNet & 80.8±0.3 & 62.3±0.5 & 81.4±0.4 & 68.1±0.6 & \textbf{83.2±0.3} & 74.8±0.3 & 82.9±0.6 & 79.7±0.6 & 82.7±0.5 & \textbf{80.7±0.3} & 82.3±0.7 & 78.5±0.3 \\
			& UNet++ & 79.9±0.1 & 64.7±0.3 & 81.5±0.4 & 73.3±0.2 & 81.6±0.6 & 72.5±0.2 & 82.1±0.3 & \textbf{78.4±0.2} & \textbf{83.6±0.2} & 77.1±0.4 & 83.0±0.4 & 75.3±0.4 \\
			& ResUNet & 82.5±0.6 & 73.7±0.5 & 83.1±0.6 & 75.8±0.6 & 84.0±0.6 & 76.5±0.9 & 83.2±0.7 & 79.0±0.3 & \textbf{84.0±0.5} & \textbf{82.8±0.3} & 83.9±0.4 & 81.4±0.1 \\
			& Attention UNet & 81.6±0.4 & 72.0±0.3 & 82.2±0.5 & 79.4±0.5 & 83.5±0.4 & 79.6±0.8 & \textbf{84.4±0.5} & 79.5±0.5 & 82.6±0.3 & 81.5±0.6 & 83.2±0.2 & \textbf{81.6±0.2} \\
			\bottomrule
	\end{tabular}}
\end{table*}

\subsubsection{Visual Turing Test}
To evaluate the perceptual realism of the images generated by our proposed method, we performed a visual Turing test involving three board-certified sonographers, each with more than 2 years of experience in BUS imaging.
For each of the five datasets, 100 images were randomly selected, including an equal number of real and synthetic images generated by our approach.
The images were randomly mixed and presented blind.
Sonographers were instructed to classify each image as real or synthetic.
Table~\ref{tab:turing} reports the classification accuracy of each sonographer, along with the mean accuracy between the raters for each dataset.
In general, the results show that our mean accuracy ranged from 46.3\% to 57.3\%, indicating that the synthetic images generated by our method are visually challenging to distinguish from the real images.

\subsection{Evaluation on Mask Generator}
We evaluated the effectiveness of the proposed SCMG through both quantitative analysis and visual inspection.

Table~\ref{tab:curvature_ablation} reports the boundary curvature values for benign and malignant tumors, with and without $\mathcal{L}_{\text{cur}}$.
The results reveal that, without $\mathcal{L}_{\text{cur}}$, the generator failed to differentiate between benign and malignant morphologies, yielding nearly identical curvature distributions.
In contrast, incorporating curvature loss largely reduced curvature in benign masks (from 0.723 to 0.619), indicating smoother boundaries, while increasing it in malignant masks (from 0.721 to 0.831), encouraging sharper and more irregular contours.

Furthermore, Fig.~\ref{fig:mask results} visualizes representative tumor masks generated under benign and malignant conditions.
It is explicitly observed that curvature constraint resulted in cleaner, elliptical masks for benign cases, while its absence led to subtle distortions.
For malignant cases, $\mathcal{L}_{\text{cur}}$ enhanced contour complexity, better capturing the spiculated, lobulated morphology typical of malignant tumors.
These results confirm that $\mathcal{L}_{\text{cur}}$ served as a structural prior, enforcing smoothness in benign tumor shapes and promoting irregularity in malignant ones, thus aligning with BI-RADS morphological descriptors.

\begin{figure}[t]
	\centering
	\includegraphics[width=\columnwidth]{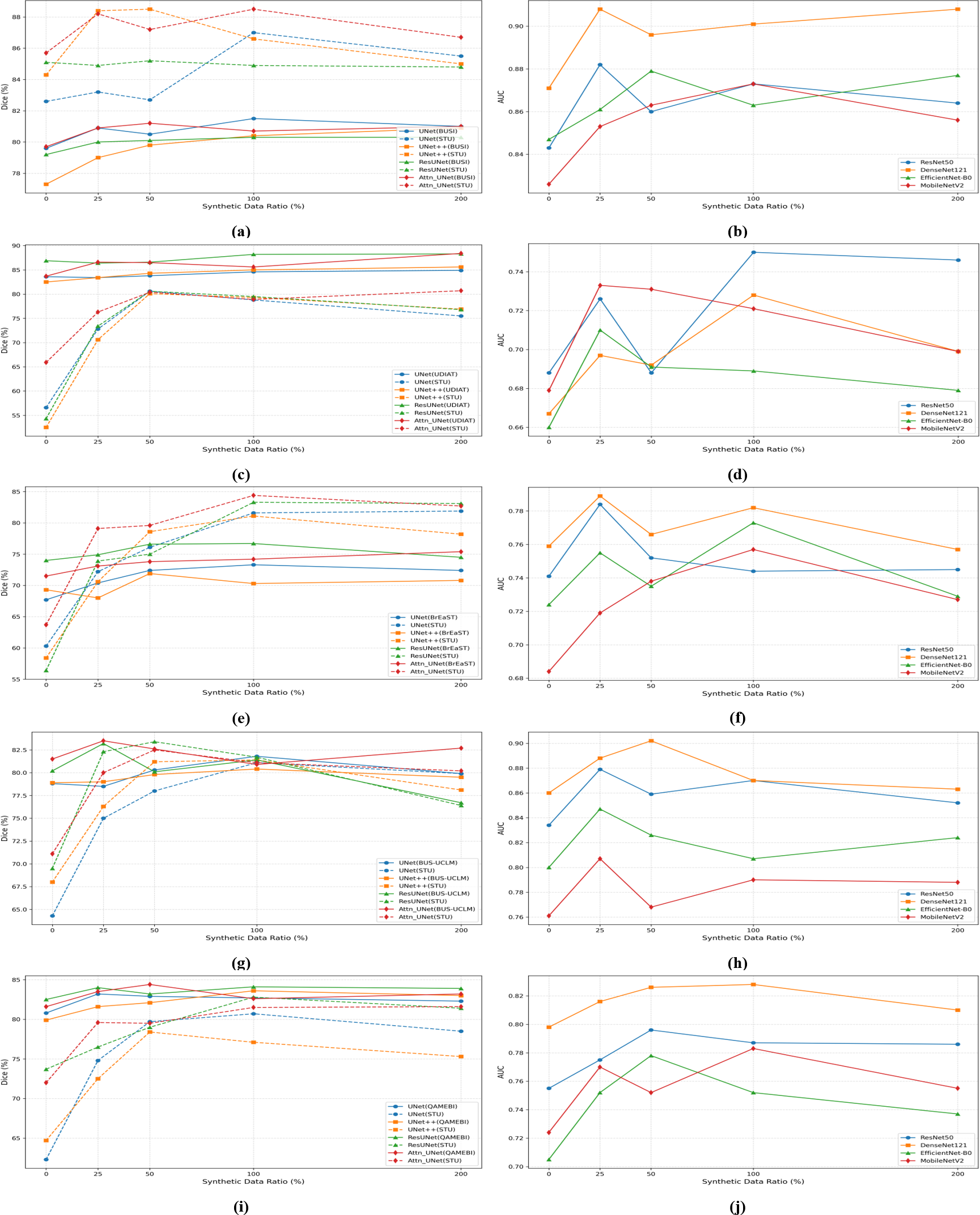}
	\caption{Downstream segmentation and classification performance under varying synthetic-to-real mixing ratios across different datasets.
		Left column (a, c, e, g, i): segmentation DSC scores on internal test sets (solid lines) and external STU set (dashed lines) across four segmentation models.
		Right column (b, d, f, h, j): classification AUC on the internal test set across four classification models.
		Rows correspond to the BUSI, UDIAT, BrEaST, BUS-UCLM, and QAMEBI datasets, respectively.}
	\label{fig:all performance}
\end{figure}

\subsection{Downstream Evaluation of Synthetic Data}
To assess the practical value of our synthetic BUS images, we conducted downstream classification and segmentation experiments across different datasets.
For each dataset, we progressively augmented the real training set with synthetic images generated by our clinically-informed pipeline at different synthetic-to-real ratios (25\%, 50\%, 100\%, 200\%), and compared these against a baseline (0\%) and ordinary augmentation strategy (combined random flipping, brightness adjustment, affine and rotation).
Classification models included ResNet50~\citep{he2016deep}, DenseNet121~\citep{huang2017densely}, EfficientNet-B0~\citep{tan2019efficientnet}, and MobileNetV2~\citep{sandler2018mobilenetv2};
segmentation models included UNet~\citep{ronneberger2015u}, UNet++~\citep{zhou2018unet++}, ResUNet~\citep{diakogiannis2020resunet}, and Attention UNet~\citep{oktay2018attention}.
For the segmentation task, each model was evaluated on the original testing dataset (internal) and the STU dataset (external).

\subsubsection{Performance Gains Across Tasks and Architectures}
As shown in Table~\ref{tab:seg-summary}, Table~\ref{tab:cls-summary} and Fig.~\ref{fig:all performance}, incorporating synthetic images notably boosted performance in both classification (AUC/F1-score) and segmentation (DSC\%) tasks across all datasets and architectures.
For example, on the BUSI dataset, DenseNet121 achieved an AUC of 0.908 at 25\% synthetic ratio (vs. 0.871 at baseline), and UNet reached 81.5\% DSC on internal data with 100\% synthetic images (vs. 79.6\% at baseline).
Similar trends can be observed on other datasets, indicating consistent enhancement due to synthetic data.

\subsubsection{Comparison with Ordinary Augmentation}
Ordinary augmentation techniques employed in our experiments included random flipping, rotation, affine transformations, and brightness adjustments.
As shown in Table~\ref{tab:cls-summary} and Table~\ref{tab:seg-summary},
except for the only case of BUSI-STU segmentation using ResUNet,
our synthetic augmentation strategy outperformed traditional augmentation strategies in both tasks.
On the BrEaST dataset, for instance, ResNet50 achieved 0.784 AUC with 25\% synthetic data, exceeding both the baseline (0.741) and ordinary augmentation (0.743).
For the segmentation task, Attention UNet yielded an external DSC of 84.4\% under 100\% synthetic augmentation, obviously higher than 63.7\% without augmentation and 66.4\% under ordinary augmentation.
These results underscore the effectiveness of structure-aware synthetic samples in providing clinically controllable diversity beyond simple image perturbations.

\subsubsection{Optimal Ratios and Non-Monotonic Trends}
It can be observed from Fig.~\ref{fig:all performance} that optimal performance often arose in moderate synthetic ratios (25\%-100\%).
Higher proportions like 200\% sometimes led marginal improvement or even slight degradation.
For example, the classification task using EfficientNet-B0 on the UDIAT dataset peaked in a synthetic ratio of 25\%, with diminishing returned in higher proportions.
Segmentation tasks show more nuanced behavior:
while 200\% synthetic data achieved optimal performance on UDIAT's internal test set in all segmentation backbones, this ratio was not the best setting for the external STU dataset.
The observed nonmonotonic trend suggests a complex interaction between the real and synthetic domains,
in connection with different tasks, datasets, and models.

\subsubsection{Cross-Dataset Generalization}
Notably, segmentation performance on the external STU dataset improved when training included synthetic data.
This suggests that our method enhanced generalization across unseen domains. 
For example, on QAMEBI, UNet reached 80.7\% DSC on STU with 100\% synthetic data, compared to 62.3\% at baseline and 68.1\% at ordinary augmentation.
This confirms the utility of clinically controllable synthetic data in improving out-of-distribution robustness.

On the whole, all the downstream experimental results demonstrate that semantically and structurally guided synthetic data can substantially enhance classification and segmentation performance beyond conventional augmentation.
In particular, optimal performance was often achieved within a moderate synthetic-to-real mixing range, rather than requiring large-scale synthetic datasets.
This suggests that effective gains can be obtained even at relatively low mixing ratios, which is particularly beneficial in resource-constrained settings.

\section{Conclusion}
In this study, we propose a clinically controllable framework for synthesizing high-fidelity breast ultrasound images under the guidance of both text prompts and tumor masks.
Integrating clinical priors, including BI-RADS-aligned descriptions and shape-aware mask synthesis, ensures synthetic images reflect diagnostically relevant features.
The dedicated mask generator enables flexible control over tumor morphology and location, while a curvature-guided loss enforces realistic boundary characteristics.
Comparison experiments across different datasets, along with ablation studies and visual Turing tests demonstrate the efficacy of our generative method.
In addition, extensive downstream experiments demonstrate that the incorporation of synthetic data effectively enhances classification and segmentation performance across multiple model architectures, even with limited augmentation ratios.
Importantly, our analysis reveals a saturation point beyond which additional synthetic data yield diminishing returns, underscoring the need for quality-focused over quantity-driven augmentation strategies.
These results highlight the potential of our framework to support clinically meaningful image generation and improve the training of CAD systems in data-scarce environments.


\bibliographystyle{model2-names.bst}\biboptions{authoryear}
\bibliography{ref}


\end{document}